\documentclass[epj]{svjour}
\usepackage{subfigure}
\usepackage{amssymb,amsmath}
\usepackage{CJK}
\usepackage{indentfirst}
\usepackage{amsmath}
\usepackage[square, comma, sort&compress, numbers]{natbib}
\usepackage{lineno,hyperref}
\usepackage{graphics}
\usepackage{graphicx}
\usepackage{multirow}
\usepackage{breqn}
\usepackage{array}
\begin{document}
\title{Thermodynamics of de Sitter black holes in massive gravity}
\author{Yu-Bo Ma\inst{1,2}, \ Ren Zhao\inst{1,2}, Shuo Cao\inst{3}
\thanks{\emph{e-mail:} caoshuo@bnu.edu.cn}%
}                     
%
%
\institute{ School of Physics, Shanxi Datong University, Datong, 037009, China \and Institute of Theoretical Physics, Shanxi Datong University, Datong 037009, China \and Department of Astronomy, Beijing Normal University, Beijing, 100875, China}
%
%
\abstract{ In this paper, by taking de Sitter space-time as a thermodynamic
system, we study the effective thermodynamic quantities of de Sitter
black holes in massive gravity, and furthermore obtain the effective
thermodynamic quantities of the space-time. Our results show that
the entropy of this type of space-time takes the same form as that
in Reissner-Nordstr\"{o}m-de Sitter space-time, which lays a solid
foundation for deeply understanding the universal thermodynamic
characteristics of de Sitter space-time in the future. Moreover, our
analysis indicates that the effective thermodynamic quantities and
relevant parameters play a very important role in the investigation
of the stability and evolution of de Sitter space-time.
} 
\maketitle
\section{Introduction}\label{sec:Intro}

The study of the thermodynamic characteristics of de Sitter space-time has arouse extensive attention in the recent years \cite{1,2,3,4,5,6,7,8,9,10,11,12,13,14,15,16,17,18,19,20,21}. At the stage of cosmological inflation in the early time, our universe behaves like a quasi-de Sitter space-time, in which the cosmological constant takes the form of vacuum energy. Moreover, if the dark energy is simply a cosmological constant, i.e., a component with constant equation of state, our universe will evolve into a new stage of de Sitter space-time in this simplest scenario. Therefore, a better knowledge of de Sitter space-time (especially its classical and quantum characteristics) is very important to construct the general framework of cosmic evolution. In the previous works, the black hole horizon and the cosmological horizon are always treated as two independent thermodynamic systems \cite{4,5,6,7,13}, from which the thermodynamic volume of de Sitter space-time as well as the corresponding thermodynamic quantities satisfying the first thermodynamics law were obtained \cite{3}. It is commonly recognized that the entropy of de Sitter space-time is the sum of that for the two types of horizons \cite{7,14}, however, such statement concerning the nature of de Sitter space-time entropy still needs to be checked with adequate physical explanation.

Considering the fact that all thermodynamic quantities related to the black hole horizon and the cosmological horizon in de Sitter space-time can be expressed as a function of mass $M$, electric charge $Q$, and cosmological constant $\Lambda$ , it is natural to consider the dependency between the two types of thermodynamic quantities. More specifically, the discussion of the following two problems is very significant to study the stability and evolution of de Sitter space-time: Do the thermodynamic quantities follow the behavior of their counterparts in AdS black holes, especially when the black hole horizon is correlated with that of the cosmological horizon? What is the specific relation between the entropy of de
Sitter space-time and that of the two horizons (the black hole horizon and the cosmological horizon)? The above two problems also provide the main motivation of this paper.

Following this direction, in our analysis we obtain the effective thermodynamic quantities of de Sitter black holes in massive gravity (DSBHMG), based on the correlation between the black hole horizon and the cosmological horizon. Our results show that the entropy of this type of space-time takes the same form as that in Reissner-Nordstrom de Sitter space-time, which lays a solid foundation for deeply understanding the universal thermodynamic characteristics of de Sitter space-time in the future. This paper is organized as follows. In Sec.II, we briefly introduce the thermodynamic quantities of the horizons of black holes and the Universe in DSBHMG, and furthermore obtain the electric charge $Q$ when the two horizons show the same radioactive temperature. In Sec.III, taking the correlation between the two horizons into consideration, we will present the equivalent thermodynamic quantities of DSBHMG satisfying the first thermodynamic law, and
perform a quantitative analysis of the corresponding effective temperature and pressure. Finally, the main conclusions are summarized in Sec.IV. Throughout the paper we use the units $G=\hbar=k_{B}=c=1$ .

\section{Thermodynamics of black holes in massive gravity}
\label{sec:1}
In the framework of $(3+1)$-dimensional massive gravity with
a Maxwell field (denoting $F_{\mu \nu } $ as the Maxwell
field-strength tensor), the corresponding action always expresses as
\cite{32,33,34,43},
\begin{equation}
\label{eq1} S=\frac{1}{2k^2}\int {d^4x\sqrt g } \left[
{R-2\Lambda-\frac{1}{4}F^2+m^2\sum\limits_i^4 {c_i \mu _i } }
\right],
\end{equation}
Here $k=+1,0,-1$ respectively correspond to the sphere, plane, and
hyperbola symmetric cases; $\Lambda $ is the cosmological constant
and $\mu _i $ represents the contribution of the matrix $\sqrt
{g^{\mu \alpha }f_{\alpha \nu}}$ with fixed symmetric tensor $f_{\mu
\nu}$. Therefore, generated form the above action, the space-time
metric of static black holes (denoting $h_{ij}$ as Einstein space
with constant curvature) can be written as
\begin{equation}
\label{eq2} ds^2=-f(r)dt^2+f^{-1}dr^2+r^2h_{ij} dx^idx^j, \quad
i,j=1,2
\end{equation}
with the metric function expressed as \cite{35,36}
\begin{equation}
\label{eq3} f(r)=k-\frac{\Lambda }{3}r^2-\frac{m_0
}{r}+\frac{q^2}{4r^2}+\frac{c_1 m^2}{2}r+m^2c_2
\end{equation}
Note that the positions of black hole horizon $r_+ $ and cosmic
horizon $r_c$ are determined when $f(r_{+,C})=0$.

\begin{center}
 \begin{figure}[htbp]
\centering
 \includegraphics[width=0.35\textwidth]{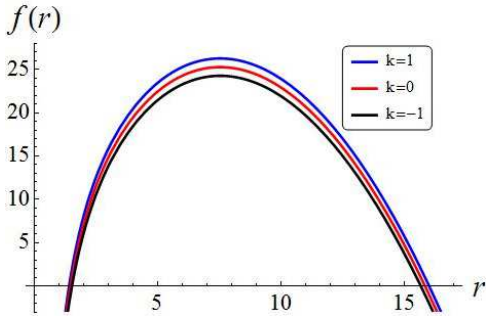}
 \caption{\label{fig:1} The metric function $f(r)$ varying with $r$.}
 \end{figure}
 \end{center}
In Fig.~1 we display the behavior of the metric function $f(r)$,
where the parameters are chosen as $\Lambda=1,m_0=30,m=2.12,c_1
=2,c_2=3.18,q=1.7$, while $k$ is fixed at $1,0$, and $-1$. It is
obvious that there are two intersection points between $f(r)$ and
the axis of $r$, which respectively correspond to the positions of
black hole horizon $r_+ $ and cosmological horizon $r_c $. Thus, the
mass $m_0$ can be expressed in terms of $r_{+,c}$ as
\begin{equation}
\begin{aligned}
\label{eq4} M=\frac{m_0 }{2} & =\frac{(k+m^2c_2 )r_c x(1+x)}{2(1+x+x^2)}\\
              &+\frac{q^2(1+x)(1+x^2)}{8r_c x(1+x+x^2)}+\frac{r_c^2 m^2c_1 x^2}{4(1+x+x^2)}
\end{aligned}
\end{equation}
and
\begin{equation}
\label{eq5} \frac{\Lambda }{3}r_c^2 (1+x+x^2)=k-\frac{q^2}{4r_c^2
x}+\frac{c_1 }{2}m^2r_c (1+x)+m^2c_2
\end{equation}

where $x=r_+ /r_c$. The temperature of the black hole horizons and
cosmic horizon can be written as \cite{37}
\begin{equation}
\begin{aligned}
&\label{eq6} T_{+,c}=\pm \frac{f(r_{+,c} )}{4\pi }\\
&=\frac{1}{4\pi r_{+,c} }\left[k-\Lambda r_{+,c}^2 -\frac{q^2}{4r_{+,c}^2 }+m^2 c_1 r_{+,c}+m^2 c_2\right]
\end{aligned}
\end{equation}
Turning to the contribution of the electrical charge $q$, it
will also generate a chemical potential as
\begin{equation}
\label{eq7} \mu _{+,c} =\frac{q}{r_{+,c} }.
\end{equation}
According to the Hamiltonian approach, we have the mass $M$ and
electric charge $Q$ as
\begin{equation}
\label{eq8} M=\frac{\nu _2 m_0 }{8\pi }, \quad Q=\frac{\nu _2
q}{4\pi }.
\end{equation}
and the entropy of the two horizons respectively express as
\begin{equation}
\label{eq9} S_{+,c} =\frac{\nu _2 }{4}r_{+,c}^2
\end{equation}
where $\nu _2 $ is the area of a unit volume of constant $(t,r)$
space (which equals to $4\pi $ for $k=0)$.
It is apparent that the thermodynamic quantities corresponding to
the two horizons satisfy the first law of thermodynamics
\begin{equation}
\label{eq10} dM=T_{+,c} dS_{+,c} +V_{+,c} dP+\mu _{+,c} dQ
\end{equation}
where
\begin{equation}
\label{eq11} V_{+,c} =\frac{\nu _2 }{3}r_{+,c}^3 , \quad
P=-\frac{\Lambda }{8\pi }
\end{equation}
When the temperature of the black hole horizon is equal to that of
the cosmological horizon, the electric charge $Q$ and the
cosmological constant $\Lambda $ are related as
\begin{equation}
\begin{aligned}
&\label{eq12} \frac{1}{r_+ }\left[ {k-\Lambda } \right.r_+^2
-\frac{q^2}{4r_+^2 }+m^2c_1 r_+ +m^2c_2 ]\\
=&-\frac{1}{r_c }\left[
{k-\Lambda } \right.r_c^2 -\frac{q^2}{4r_c^2 }+m^2c_1 r_c +m^2c_2 ]
\end{aligned}
\end{equation}
As can be seen from Eq. (\ref{eq5}) and (\ref{eq12}) the electric
charge of the system satisfies the following expression
\begin{equation}
\label{eq13} \frac{q^2(1+x)^2}{4r_c^2 x^2}=k+\frac{c_1 m^2r_c
x}{2(1+x)}+m^2c_2 .
\end{equation}
When taking $T_+ =T_c $, the combination of Eqs. (\ref{eq5}),
(\ref{eq6}) and (\ref{eq13}) will lead to the temperature $T$ as
\begin{equation}
\begin{aligned}
\label{eq14}&T=T_+ =T_c \\
=&\frac{(1-x)}{2\pi r_c (1+x)^2}\left[
{k+\frac{m^2c_1 r_c (1+4x+x^2)}{4(1+x)}+m^2c_2 } \right]
\end{aligned}
\end{equation}
\section{Effective thermodynamic quantities}\label{sec:2}
Considering the connection between the black hole horizon
and the cosmological horizon, we can derive the effective
thermodynamic quantities and corresponding first law of black hole
thermodynamics as
\begin{equation}
\label{eq15} dM=T_{eff} dS-P_{eff} dV+\phi _{eff} dQ,
\end{equation}
where the thermodynamic volume is defined by \cite{3,5,6,38}
\begin{equation}
\label{eq16} V=\frac{4\pi }{3}\left( {r_c^3 -r_+^3 } \right).
\end{equation}
It is obvious that there exit three real roots for the equation
$f(r)=0$: the cosmological horizon (CEH) $r=r_c $, the inner
(Cauchy) horizon of black holes, and the outer horizon (BEH) $r=r_+
$ of black holes. Moreover, the de Sitter space-time is
characterized by $\Lambda >0$, while $\Lambda <0$ denotes the
anti-de Sitter scenario.

Now we will consider the interaction between the black hole horizon
and the cosmological horizon \cite{39,40}
\begin{equation}
\label{eq17} S=\pi br_c^2 \left[1+x^2+f(x)\right],
\end{equation}
Here the undefined function $f(x)$ represents the extra contribution
from the correlations of the two horizons. From Eq. (\ref{eq15}), we
can obtain the effective temperature $T_{eff}$ and pressure $P_{eff}
$
\begin{equation}
\begin{aligned}
\label{eq18} T_{eff}& =\left( {\frac{\partial M}{\partial S}}\right)_{Q,V}\\
& =\frac{\left( {\frac{\partial M}{\partial x}}
\right)_{r_c } \left( {\frac{\partial V}{\partial r_c }} \right)_x
-\left( {\frac{\partial V}{\partial x}} \right)_{r_c } \left(
{\frac{\partial M}{\partial r_c }} \right)_x }{\left(
{\frac{\partial S}{\partial x}} \right)_{r_c } \left(
{\frac{\partial V}{\partial r_c }} \right)_x -\left( {\frac{\partial
V}{\partial x}} \right)_{r_c } \left( {\frac{\partial S}{\partial
r_c }} \right)_x },
\end{aligned}
\end{equation}
\begin{equation}
\begin{aligned}
\label{eq19} P_{eff} &=-\left( {\frac{\partial M}{\partial V}}\right)_{Q,S}\\
 &=-\frac{\left( {\frac{\partial M}{\partial x}}
\right)_{r_c } \left( {\frac{\partial S}{\partial r_c }} \right)_x
-\left( {\frac{\partial S}{\partial x}} \right)_{r_c } \left(
{\frac{\partial M}{\partial r_c }} \right)_x }{\left(
{\frac{\partial V}{\partial x}} \right)_{r_c } \left(
{\frac{\partial S}{\partial r_c }} \right)_x -\left( {\frac{\partial
S}{\partial x}} \right)_{r_c } \left( {\frac{\partial V}{\partial
r_c }} \right)_x }.
\end{aligned}
\end{equation}
Combining Eqs.(\ref{eq4}), (\ref{eq16}) and (\ref{eq17}), one can
obtain
\begin{equation}
\label{eq20} T_{eff} =\frac{B(x,q)}{2\pi r_c \left[
{2x^2(1+x^2+f(x))+(1-x^3)(2x+f'(x))} \right]},
\end{equation}
where
\begin{equation}
\begin{aligned}
\label{eq21}
B(x,q)=&(k+m^2c_2 )\frac{(1+x-2x^2+x^3+x^4)}{(1+x+x^2)}\\
&-\frac{q^2}{4r_c^2 }\frac{(1+x+x^2-2x^3+x^4+x^5+x^6)}{x^2(1+x+x^2)}\\
&+r_c m^2c_1 x\frac{(2+x)(1-x)+2x^3}{2(1+x+x^2)}.
\end{aligned}
\end{equation}
\begin{equation}
\label{eq22} P_{eff} =\frac{D(x,q)}{8\pi r_c^2
\left[{2x^2(1+x^2+f(x))+(1-x^3)(2x+f'(x))} \right]},
\end{equation}

\begin{equation}
\label{eq23}
\begin{aligned}
D(x,q)=&\frac{(k+m^2 c_2)}{(1+x+x^2)^2}\left\{\left((1+2x)(1+x^2+f(x))\right)\right.\\
       &\left.-x(1+x)(1+x+x^2)(2x+f'(x))\right\}\\
       &-\frac{q^2}{4r_c^2x^2(1+x+x^2)^2}\left\{\left(2(1+2x+3x^2)\right)\right.\\
       &(1+x^2+f(x))-x(1+x)(1+x^2)(1+x+x^2)\\
       &\left.\left.(2x+f'(x))\right) \right\}+\frac{r_c m^2c_1 x}{(1+x+x^2)^2}\left\{\left(2(2+x)\right. \right. \\
       &\left.\left.(1+x^2+f(x))-x(1+x+x^2)(2x+f'(x))\right)\right\},
\end{aligned}
\end{equation}

When the temperature of the black hole horizon is equal to that of
the cosmological horizon, the effective temperature of the
space-time should be
\begin{equation}
\label{eq24} T_{eff}=T=T_c =T_+ .
\end{equation}
Then substituting Eq. (\ref{eq14}) into Eq. (\ref{eq20}), we get
\begin{equation}
\label{eq25}
\begin{aligned}
&\frac{(1-x)}{(1+x)^2}\left[ {k+m^2c_2 +\frac{m^2c_1 r_c
(1+4x+x^2)}{4(1+x)}}
\right](1+x+x^2)\\
=&\frac{B(x)}{\left[ {2x^2(1+x^2+f(x))+(1-x^3)(2x+f'(x))} \right]}.
\end{aligned}
\end{equation}
where
\begin{equation}
\label{eq26} B(x)=\frac{2x(1+x^4)}{(1+x)^2}\left[ {(k+m^2c_2
)+\frac{r_c m^2c_1 }{4(1+x)}(1+4x+x^2)} \right].
\end{equation}
Then Eq. (\ref{eq25}) will transform into
\begin{equation}
\label{eq27}
f'(x)+\frac{2x^2}{1-x^3}f(x)=\frac{2x^2(2x^3+x^2-1)}{(1-x^3)^2}.
\end{equation}
with the corresponding solution as
\begin{equation}
\label{eq28} f(x)=-\frac{2\left( {4-5x^3-x^5} \right)}{5\left(
{1-x^3} \right)}+C_1 \left( {1-x^3} \right)^{2/3}.
\end{equation}
Considering the initial condition of $f(0)=0$, we can obtain $C_1
=8/5$ and inserting Eq.(\ref{eq25}) into Eqs.(\ref{eq20}) and
(\ref{eq22}) will lead to
\begin{equation}
\label{eq29} T_{eff} =\frac{B(x,q)(1-x^3)}{4\pi r_c x(1+x^4)}, \quad
P_{eff} =\frac{D(x,q)(1-x^3)}{16\pi r_c^2 x(1+x^4)}.
\end{equation}
Based on the above equations, the $P_{eff}-x$ and $T_{eff}-x$
diagrams could be derived by taking different value of $k$, $q$,
$m$, $c_1$ and $c_2 $ (when taking $r_c =1)$.

 \begin{center}
 \begin{figure}[htbp]
\centering
 \includegraphics[width=0.35\textwidth]{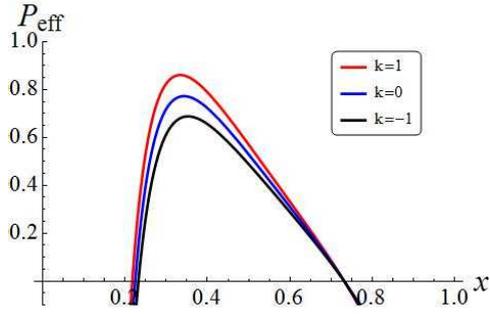}
\caption{\label{fig:2} The $P_{eff} -x$ diagram when the parameter
$k$ is fixed at 1, 0 and -1, respectively. The other parameters are
fixed at $m=2.12,c_1 =2,c_2 =3.18,q=1.7$.}
 \end{figure}
 \end{center}

 \begin{center}
 \begin{figure}[htbp]
\centering
 \includegraphics[width=0.24\textwidth]{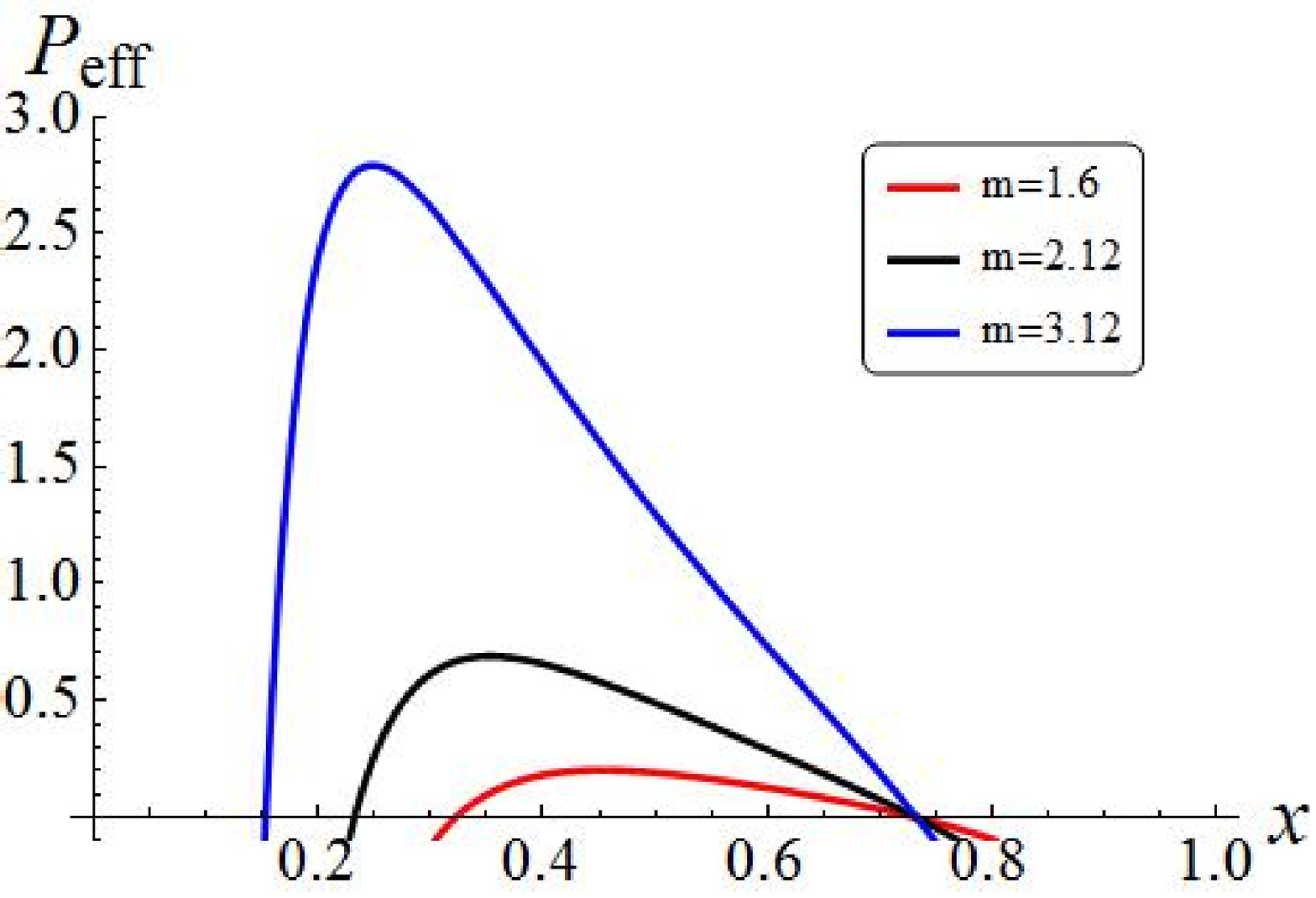} \includegraphics[width=0.24\textwidth]{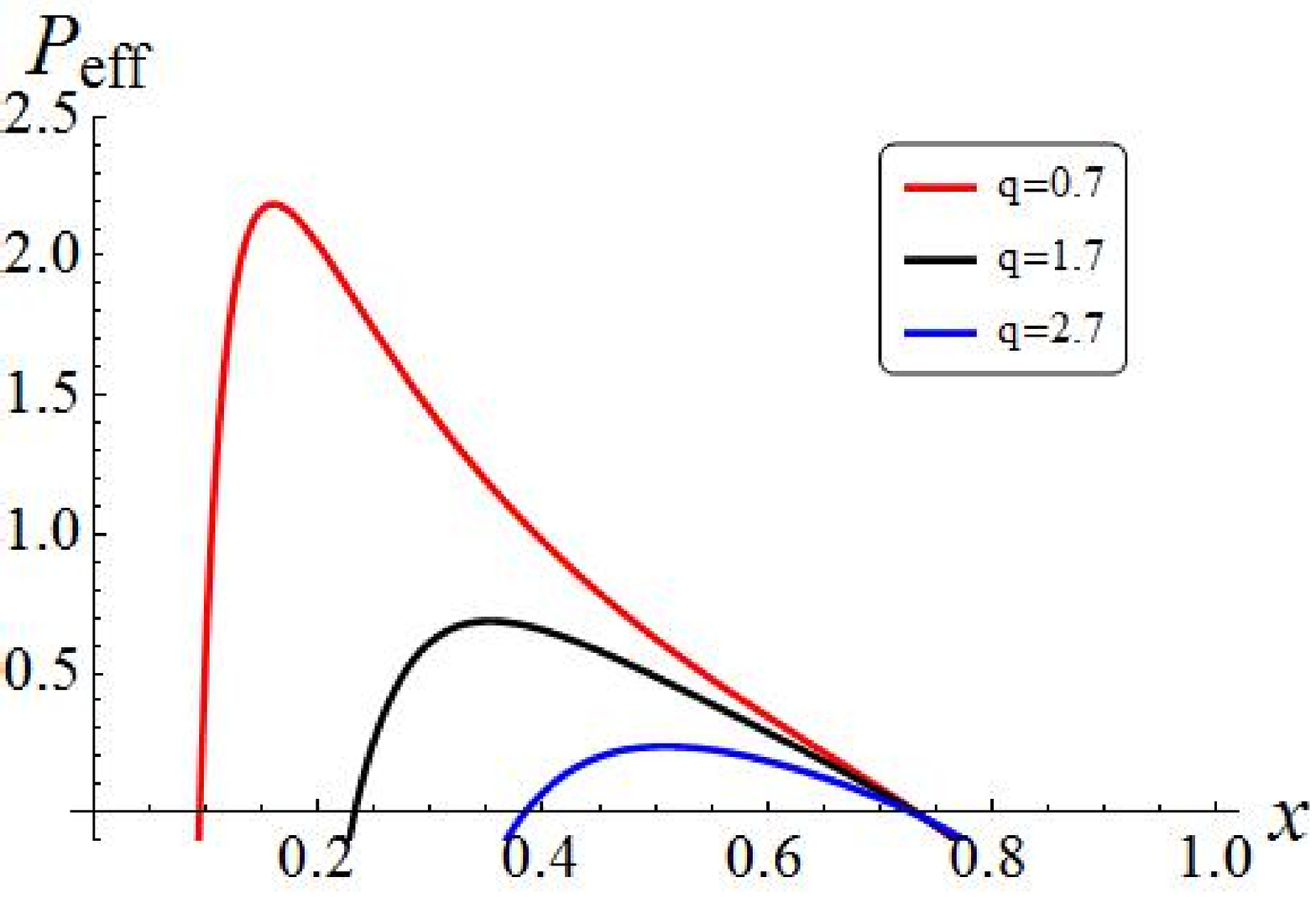}
 \includegraphics[width=0.24\textwidth]{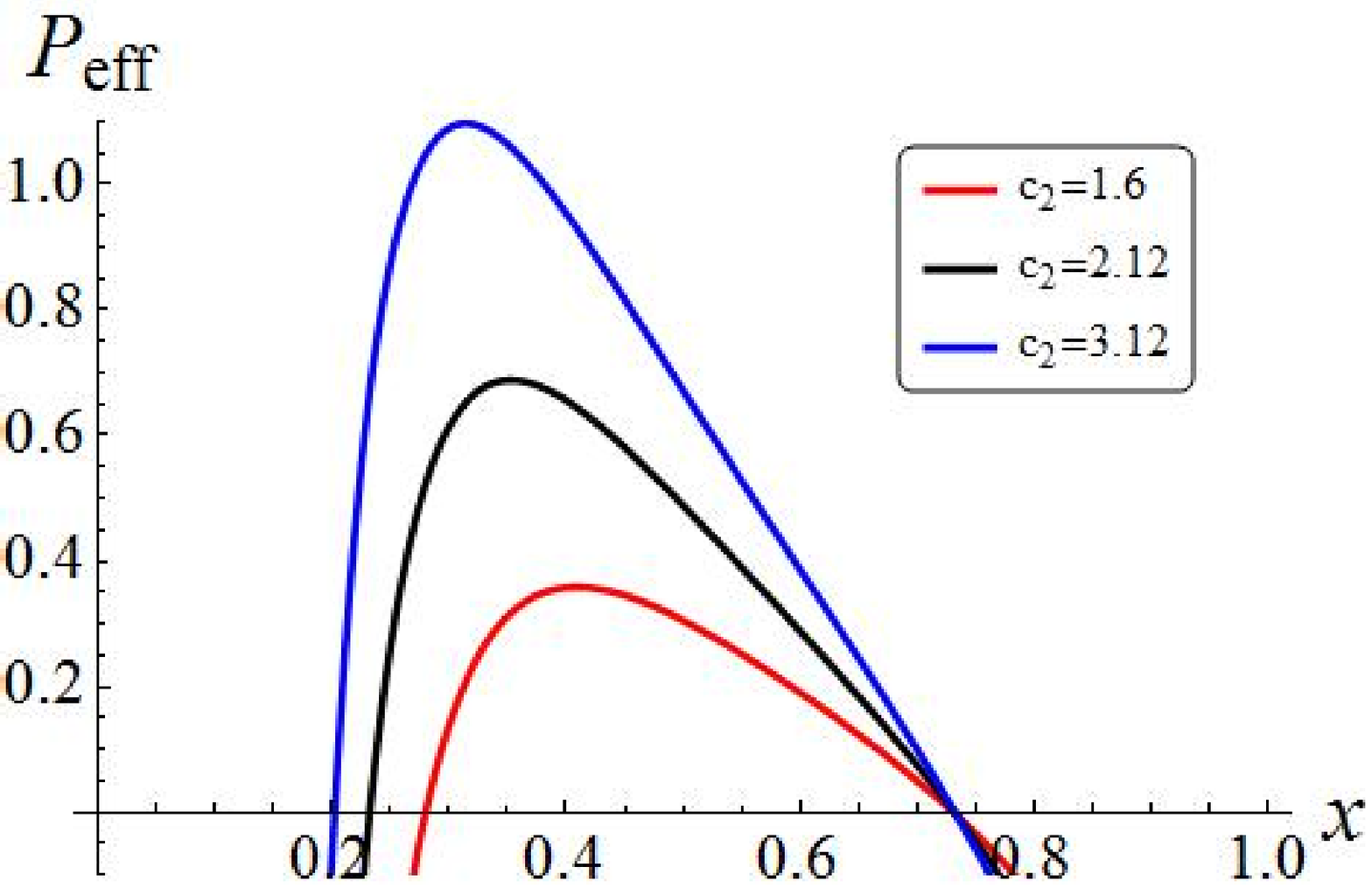} \includegraphics[width=0.24\textwidth]{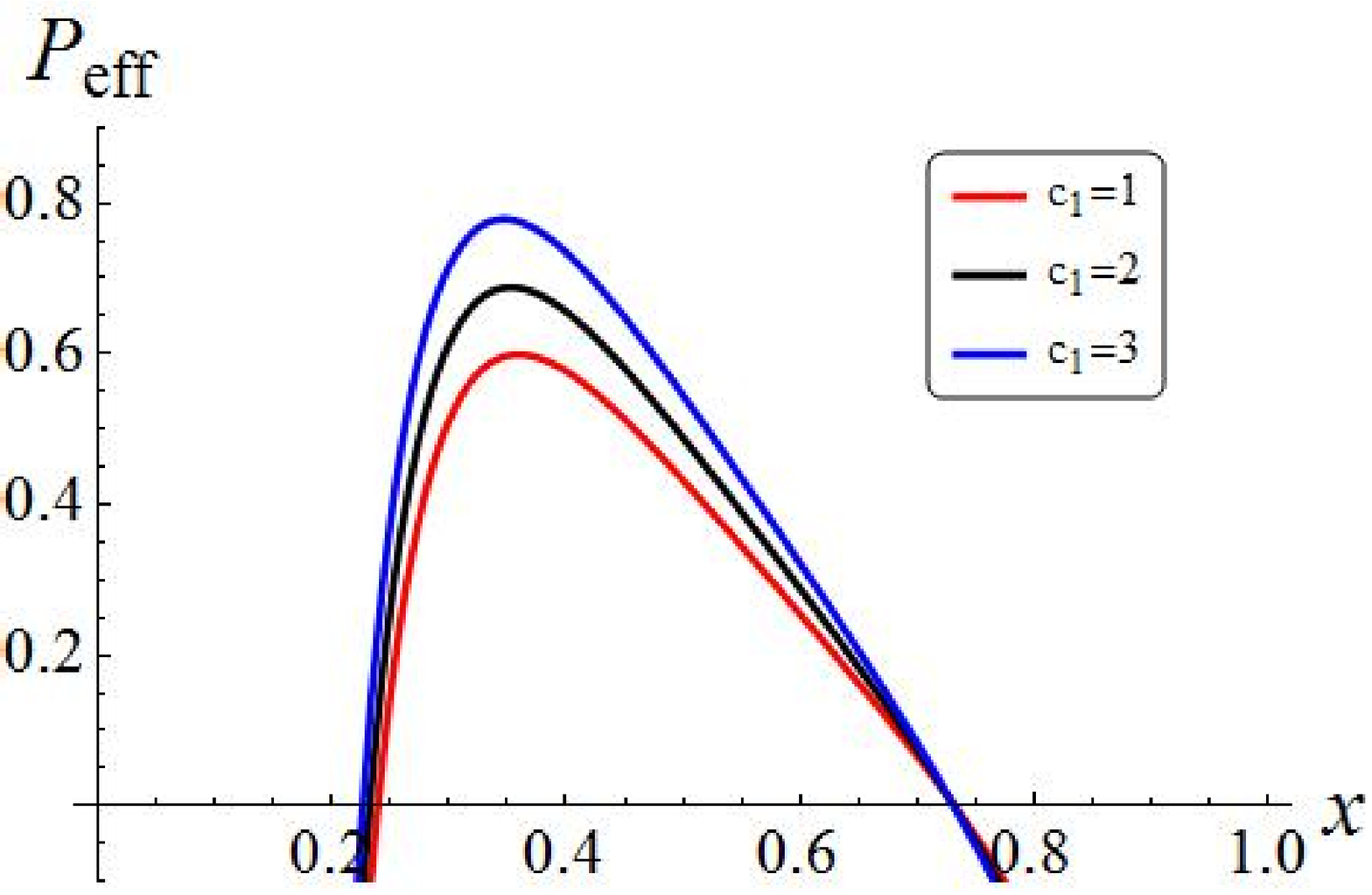}
 \caption{\label{fig:3} The $P_{eff} -x$ diagram varying with the parameters of
$m$, $c_1$, $c_2$ and $q$, while the other two parameters are fixed
at $k=-1$, $r_{c}=1$.}
 \end{figure}
 \end{center}

\begin{center}
 \begin{figure}[htbp]
\centering
 \includegraphics[width=0.24\textwidth]{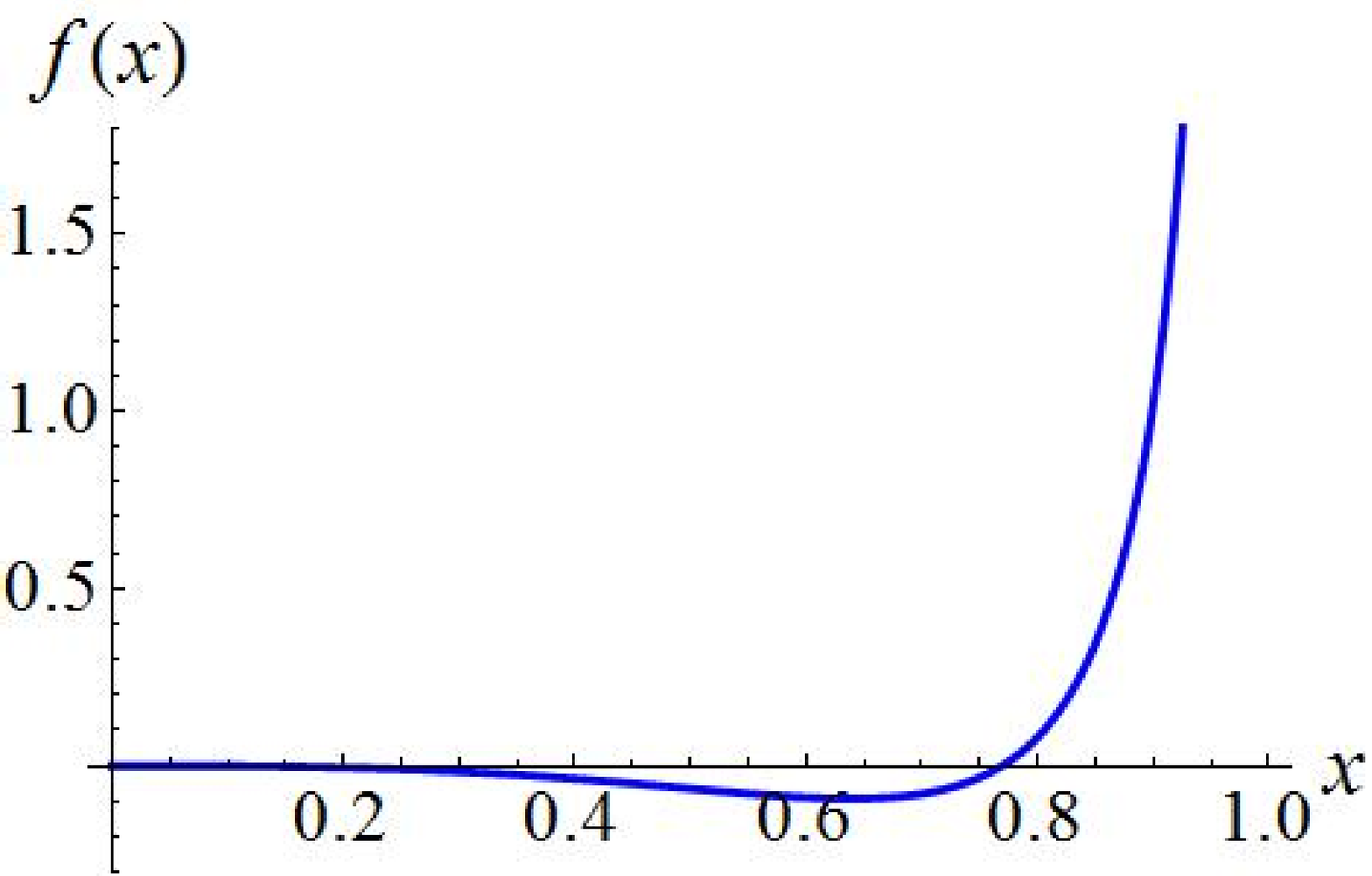} \includegraphics[width=0.22\textwidth]{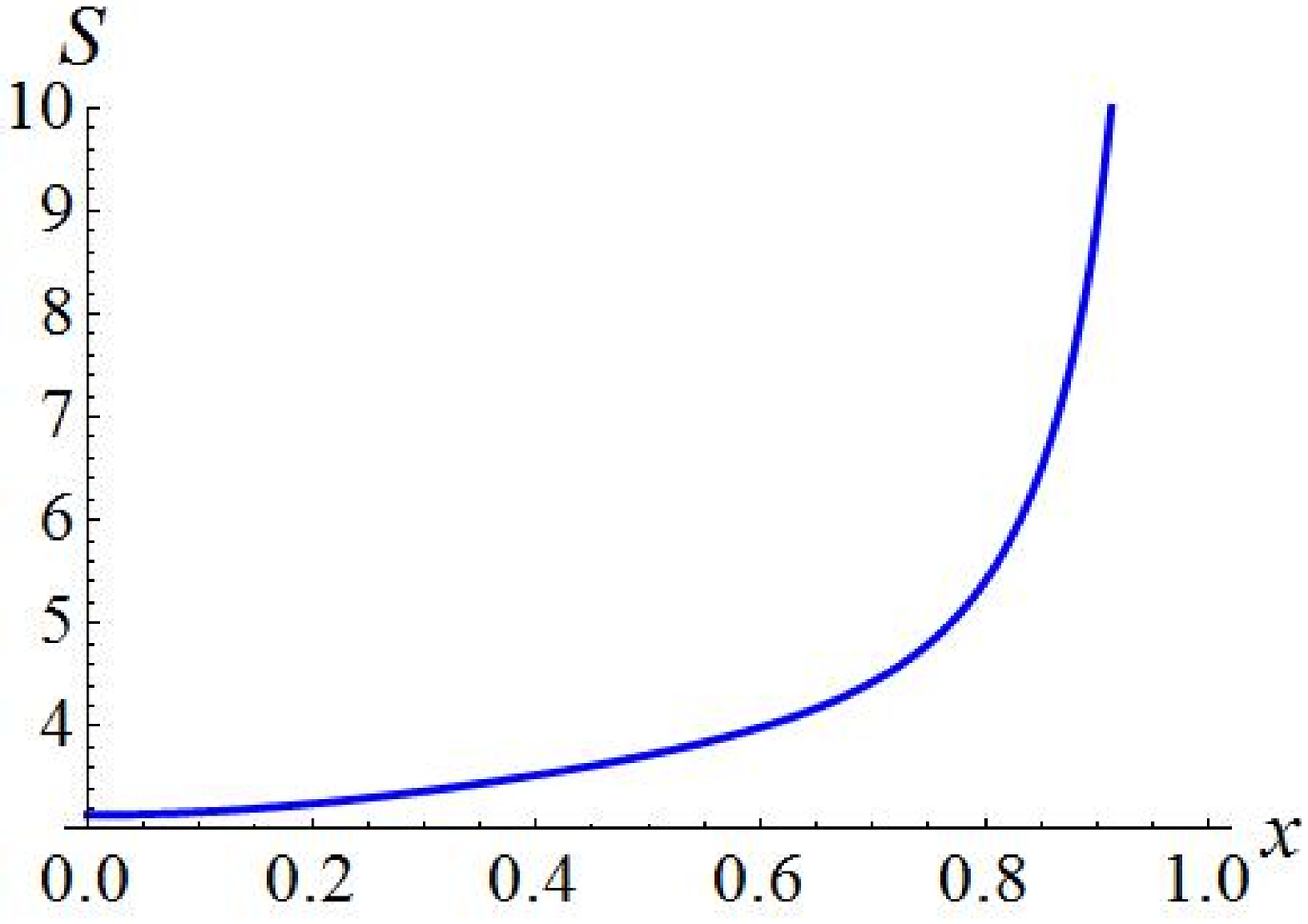}
 \caption{\label{fig:4} The $S(x)-x$ and $f(x)-x$ diagrams with $r_c =1$.}
 \end{figure}
 \end{center}
\begin{center}
 \begin{figure}[htbp]
\centering
 \includegraphics[width=0.35\textwidth]{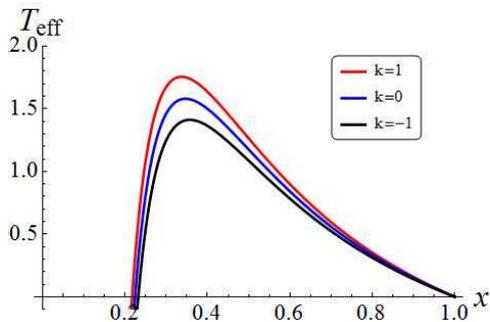}
 \caption{\label{fig:5} The $T_{eff} -x$ diagram when $k$ is fixed at 1, 0 and -1,
respectively. The other parameters are fixed at $m=2.12, c_1 =2, c_2
=3.18, q=1.7$, $r_{c}=1$.}
 \end{figure}
 \end{center}
In Fig.~2 and 3, we illustrate an example of the $P_{eff} -x$
diagram with different value of relevant parameters, from which one
could clearly see the effect of these parameters on the effective
pressure of RN-dSQ space-time. Following the same procedure by
inserting Eq. (\ref{eq28}) into Eq. (\ref{eq17}), we can also obtain
the $S(x)-x$ and $f(x)-x$ diagrams with $r_c =1$, which are
explicitly shown in Fig.~4. Similarly, in Fig. 5 and 6, we show the
evolution of the $T_{eff}-x$ diagram with different value of
relevant parameters, from which one could perceive the effect of
these parameters on the effective temperature of RN-dSQ space-time.
\begin{center}
\begin{figure}[htbp]
\centering
\includegraphics[width=0.24\textwidth]{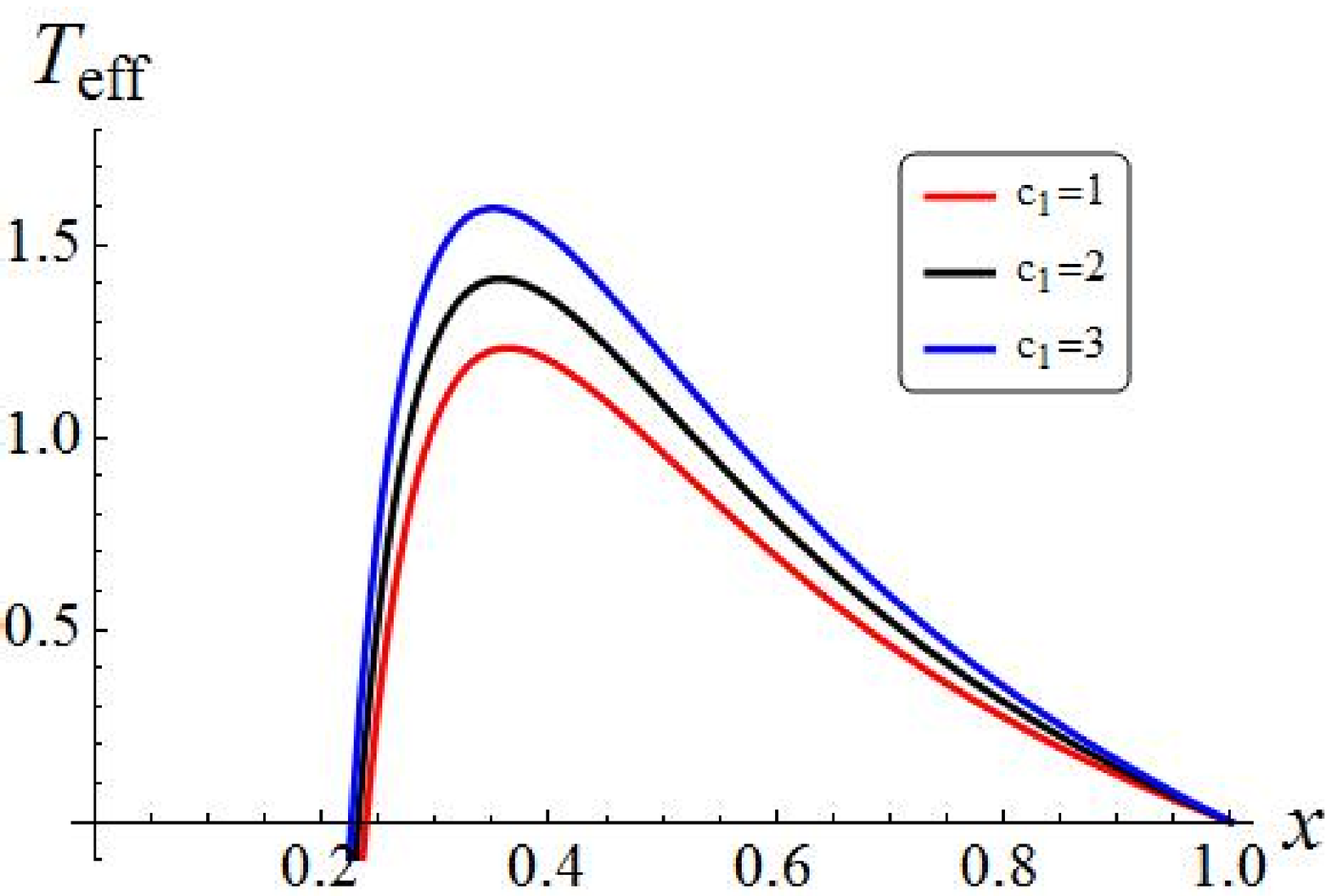} \includegraphics[width=0.24\textwidth]{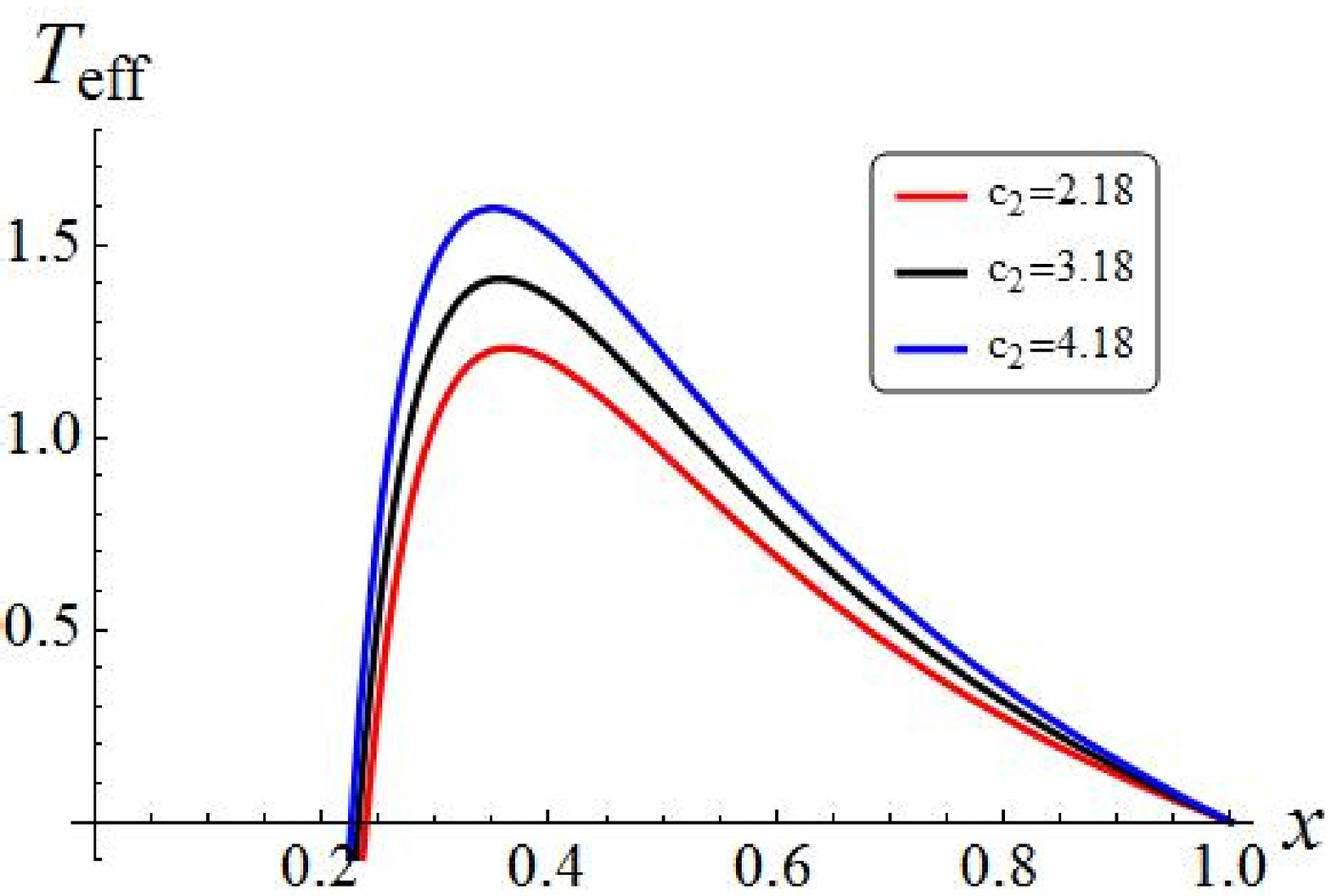}
\includegraphics[width=0.24\textwidth]{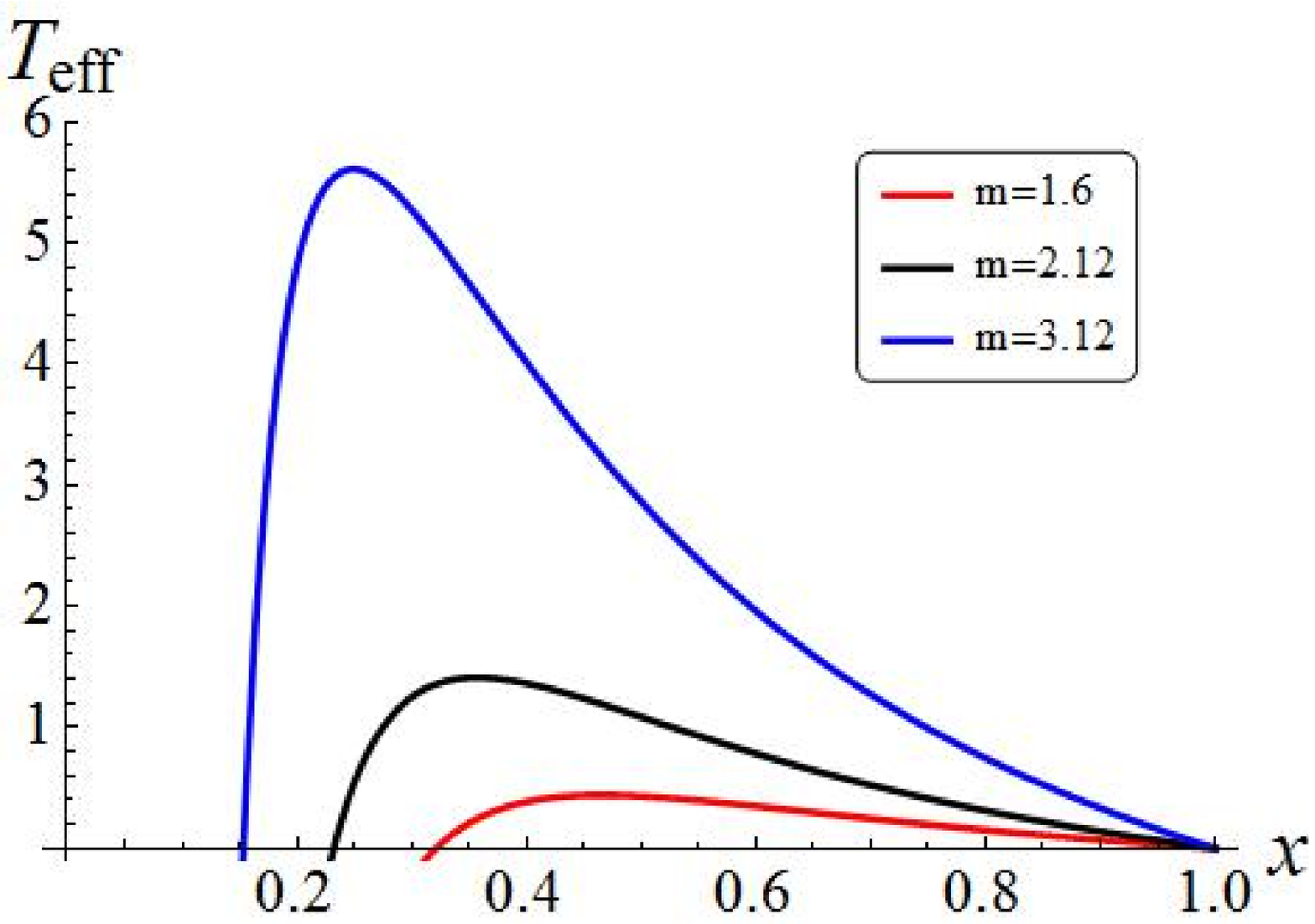} \includegraphics[width=0.24\textwidth]{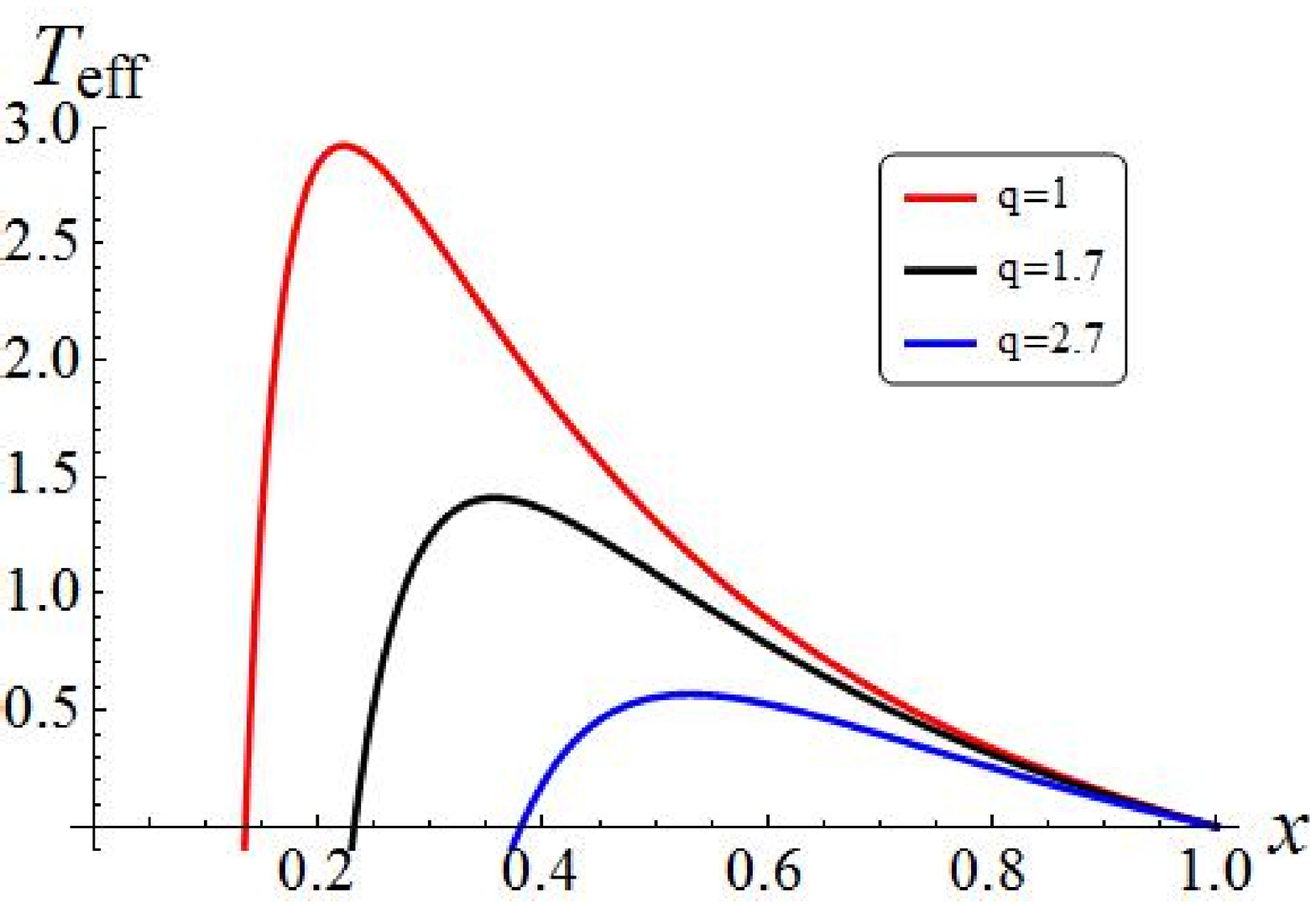}
\caption{\label{fig:6} The $T_{eff} -x$ diagram varying with the
parameters of $m$, $c_1 $, $c_2$ and $q$, while the other two
parameters are fixed at $k=-1$, $r_{c}=1$.}
\end{figure}
\end{center}
Moreover, it is shown that the change of these related parameters
may also significantly affect the position of the stability and
phase-transition points, which can be clearly seen from the results
presented in Table 1 and 2.
\begin{table}[tbhp]
 \begin{center}
 \begin{tabular}{|c|c|c|c|c|c|c|}\hline
 \cline{1-4}
 Parametric   &  $x_c $   & $T_{eff}^c $   & $x_0$ \\
        \hline
           $ k=1 $     & 0.3374    &1.7554    & 0.2173\\
        \hline
           $ k=0 $    & 0.3468    & 1.5799   & 0.2243\\
        \hline
           $ k=-1$   & 0.3571    & 1.4121    & 0.2321\\
        \hline
 \end{tabular}
 \end{center}
\caption{Summary of the highest effective temperature $T_{eff}^c $
and the corresponding $x_c $ for different curves in Fig.~5. The
value of $x_0 $ when the effective temperature reaches zero is also
listed. \label{tab:1}}
 \end{table}
\begin{table}[tbhp]
 \begin{center}
 \begin{tabular}{|c|c|c|c|c|c|c|} \hline
 \cline{1-5}
  &Parametric   &  $x_c $   & $T_{eff}^c $   & $x_0$ \\
        \hline
                        & $q=1.0$    &0.22269    & 2.92226  & 0.13498\\
           $ k=-1 $     & $q=1.7$    &0.35710    & 1.41210  & 0.23210\\
                        & $q=2.7$    &0.53108    & 0.56870  & 0.38043\\
        \hline
                        & $m=1.6$    &0.46546     & 0.43960    & 0.32108\\
           $ k=-1$      & $m=2.12$    &0.35710    &1.41210    & 0.22310\\
                        & $m=3.12$    & 0.24946   & 5.61153  & 0.15324\\
        \hline
                        & $c_1=1.0$    & 0.36373    & 1.23166  & 0.23898\\
           $ k=-1$      & $c_1=2.0$    & 0.35710    &1.41210   & 0.23210\\
                        & $c_1=3.0$    & 0.35114    & 1.59504   & 0.22599\\
        \hline
                        & $c_2=2.18$    & 0.41792    & 0.75768  & 0.27926\\
           $ k=-1$      & $c_2=3.18$    & 0.35710    & 1.41210  & 0.23210\\
                        & $c_2=4.18$    & 0.31677    & 2.22455  & 0.20218\\
        \hline
 \end{tabular}
 \end{center}
\caption{ Summary of the highest effective temperature $T_{eff}^c $
and the corresponding $x_c $ for different curves in Fig.~6. The
value of $x_0 $ when the effective temperature reaches zero is also
listed. \label{tab:2}}
 \end{table}
\section{Conclusion and discussion} \label{sec:3}

In this paper, by taking de Sitter space-time as a thermodynamic
system, we study the effective thermodynamic quantities of de Sitter
black holes in massive gravity, and furthermore obtain the effective
thermodynamic quantities of the space-time. Here we summarize our
main conclusions in more detail:

\begin{itemize}

\item In the previous analysis without considering the correlation
between the black hole horizon and the cosmological horizon, i.e.,
the two horizons are always treated as independent thermodynamic
systems with different temperature, the space-time does not satisfy
the requirement of thermodynamic stability. In this paper, we find
that the establishment of the correlation between the two horizons
will generate the common effective temperature $T_{eff}$, which may
represent the most typical thermodynamic feature of RN-dSQ
space-time.

\item As can be clearly seen from the $S(x)-x$ and $T_{eff}-x$
diagrams, RN-dSQ space-time in unstable under the condition of
$x>x_{c}$ and $x<x_{0}$. This result indicates that there exist only
the RN-dSQ black holes satisfying the condition of $x_{0}<x<x_{c}$ ,
which lays a solid theoretical foundation for the search of black
holes in the Universe.

\item We find that the interaction term $f(x)$ in the entropy of
RN-dSQ space-time takes the same form of that in RN-dS space-time.
Considering that the entropy in the two types of space-time is the
function of the position of the horizon, which has no relation with
other parameters including the electric charge $(Q)$ and the
constant $(\Lambda)$, the entropy in the two types of space-time
should take the same form. This finding may contribute to the deep
understanding the universal thermodynamic characteristics of de
Sitter space-time in the future.

\end{itemize}

\section*{Acknowledgments}

The authors declare that there is no conflict of interest regarding
the publication of this paper. This work was supported by the Young
Scientists Fund of the National Natural Science Foundation of China
(Grant Nos. 11605107 and 11503001), in part by the National Natural
Science Foundation of China (Grant No. 11475108), Supported by
Program for the Innovative Talents of Higher Learning Institutions
of Shanxi, the Natural Science Foundation of Shanxi Province, China
(Grant No.201601D102004) and the Natural Science Foundation for
Young Scientists of Shanxi Province, China (Grant No.
201601D021022), the Natural Science Foundation of Datong city (Grant
No. 20150110).

\begin{thebibliography}{00}
%
%
 \bibitem{1}
 R.G. Cai, Cardy-Verlinde formula and thermodynamics of black holes in de Sitter spaces, Nucl. Phys. B 628 (2002) 375
\bibitem{2}
 B.D. Koberlein and R. L. Mallett,  Charged, radiating black holes, inflation, and cosmic censorship, Phys. Rev. D 49 (1994) 5111
\bibitem{3}
 B.P. Dolan, D. Kastor, D. Kubiznak, R. B. Mann, J. Traschen, Thermodynamic volumes and isoperimetric inequalities for de Sitter black holes, Phys. Rev. D 87 (2013) 104017
\bibitem{4}
 Y. Sekiwa, Thermodynamics of de Sitter black holes: Thermal cosmological constant, Phys. Rev. D 73 (2006) 084009
\bibitem{5}
 D. Kubiznak, F. Simovic,  Thermodynamics of horizons: de Sitter black holes and reentrant phase transitions, Class. Quant. Grav. 33 (2016) 245001
\bibitem{6}
 J. McInerney, G. Satishchandran, J. Traschen, Cosmography of KNdS Black Holes and Isentropic Phase Transitions, Class. Quant. Grav. 33 (2016) 105007
\bibitem{7}
 M. Urano and A. Tomimatsu, H. Saida, Mechanical First Law of Black Hole Spacetimes with Cosmological Constant and Its Application to Schwarzschild-de Sitter Spacetime, Class. Quant. Grav.26 (2009) 105010
\bibitem{8}
  X.Y. Guo, H.F. Li, L.C. Zhang and R. Zhao, Thermodynamics and phase transition of in the Kerr-de Sitter black hole,  Phys. Rev. D 91 (2015) 084009
\bibitem{9}
  X.Y. Guo, H.F. Li, L.C. Zhang and R. Zhao, The critical phenomena of charged rotating de Sitter black holes, Class. Quant. Grav. 33 (2016) 135004
\bibitem{10}
  H.H. Zhao , M. S. Ma, L.C. Zhang, and R. Zhao, $P?V$ criticality of higher dimensional charged topological dilaton de Sitter black holes, Phys. Rev. D 90 (2014) 064018
\bibitem{11}
  M.S. Ma, R. Zhao, Y.Q. Ma, Thermodynamic stability of black holes surrounded by quintessence, Gen. Relativ. Gravit. 49 (2017) 79
\bibitem{12}
  T. Katsuragawa and S. Nojiri,  Stability and antievaporation of the Schwarzschild-de Sitter black holes in bigravity,  Phys. Rev. D 91 (2015) 084001
\bibitem{13}
  F. Mellor and I. Moss, Black holes and gravitational instantons,  Class. Quant. Grav. 6 (1989) 1379
\bibitem{14}
  D. Kastor and J. Traschen, Cosmological multi-black-hole solutions, Phys. Rev. D 47 (1993) 5370
\bibitem{15}
  K. Hinterbichler, Theoretical aspects of massive gravity,  Rev. Mod. Phys. 84 (2012) 671
\bibitem{16}
  M. Fierz, {\"U}ber die relativistische Theorie kr{\"a}ftefreier Teilchen mit beliebigem Spin, Helv. Phys. Acta. 12 (1939) 3-37
\bibitem{17}
  M. Fierz and W. Pauli, On relativistic wave equations for particles of arbitrary spin in an electromagnetic field, Proc. R. Soc. A 173 (1939) 211-232
\bibitem{18}
  H. V. Dam, and M. J. G. Veltman, Massive and mass-less Yang-Mills and gravitational fields,  Nucl. Phys. B 22 (1970) 397-411
\bibitem{19}
  H.F. Li, M.S. Ma, Y.Q. Ma, Thermodynamic properties of black holes in de Sitter space, Mod. Phys. Lett. A  32 (2017) 1750017
\bibitem{20}
  M. Azreg-A{\"i}nou, Black hole thermodynamics: No inconsistency via the inclusion of the missing P-V terms, Phys. Rev. D 91 (2015) 064049
\bibitem{21}
 M. Azreg-A{\"i}nou, Charged de Sitter-like black holes: quintessence-dependent enthalpy and new extreme solutions, Eur. Phys. J. C 75 (2015) 34
\bibitem{32}
  S. Upadhyay, B. Pourhassan, H. Farahani, P-V criticality of first-order entropy corrected AdS black holes in massive gravity,  Phys. Rev. D  95 (2017) 106014
\bibitem{33}
  S. F. Hassan, R. A. Rosen, On Non-Linear Actions for Massive Gravity,  JHEP 07 (2011) 009
\bibitem{34}
  A. Adams, D. A. Roberts, O. Saremi,
Hawking-Page transition in holographic massive gravity, Phys. Rev. D 91 (2015) 046003
\bibitem{43}
A. Dehyadegari, M. Kord Zangeneh, A. Sheykhi, arXiv:1703.00975
\bibitem{35}
  D.C. Zou, R.H. Yue, M. Zhang, Reentrant phase transitions of higher-dimensional AdS black holes in dRGT massive gravity, Eur. Phys. J. C  77 (2017) 256
\bibitem{36}
  P. Boonserm, T. Ngampitipan, P. Wongjun, arXiv:1705.03278
\bibitem{37}
  S.H. Hendi, R. B. Mann, S. Panahiyan, B. Eslam Panah, van der Waals like behaviour of topological AdS black holes in massive gravity, Phys. Rev. D 95 (2017) 021501
\bibitem{38}
  M. Cvetic, G.W. Gibbons, D. Kubiznak, C.N. Pope, Black hole enthalpy and an entropy inequality for the thermodynamic volume,  Phys. Rev. D 84 (2011) 024037
\bibitem{39}
  L.C. Zhang, R. Zhao, M.S. Ma, Entropy of Reissner-Nordstr{\"o}m-de Sitter black hole, Phys. Lett. B 761 (2016) 74¨C76
\bibitem{40}
  H.F. Li, M.S. Ma, L.C. Zhang, R. Zhao, Entropy of Kerr-de Sitter black hole, Nucl. Phys. B 920 (2017) 211¨C220
\end{thebibliography}
%

\end{document}